\documentclass[superscriptaddress,twocolumn,showpacs,preprintnumbers,amsmath,
amssymb,prb]{revtex4}
\usepackage{graphicx}
\usepackage{color}
\begin{document}

\title{Energetics and stability of dangling-bond silicon wires on H passivated Si(100)}

\author{Roberto Robles}
\affiliation{Centro de investigaci\'on en
  nanociencia y nanotecnolog\'{\i}a
  (CSIC - ICN), Campus de la UAB,
  E-08193 Bellaterra, Spain}
\author{Mika\"el Kepenekian}
\affiliation{Centro de investigaci\'on en
  nanociencia y nanotecnolog\'{\i}a
  (CSIC - ICN), Campus de la UAB,
  E-08193 Bellaterra, Spain}
\author{Serge Monturet}
\affiliation{Centre d'Elaboration des Mat\'eriaux et d'Etudes Structurales (CEMES)
\& MANA Satellite,
CNRS, 29 rue J. Marvig, F-31055 Toulouse, C\'edex, France}
\author{Christian Joachim}
\affiliation{Centre d'Elaboration des Mat\'eriaux et d'Etudes Structurales (CEMES)
\& MANA Satellite,
CNRS, 29 rue J. Marvig, F-31055 Toulouse, C\'edex, France}
\author{Nicol\'as Lorente}
\affiliation{Centro de investigaci\'on en
  nanociencia y nanotecnolog\'{\i}a
  (CSIC - ICN), Campus de la UAB,
  E-08193 Bellaterra, Spain}

\date{\today}

\begin{abstract}

We evaluate the electronic, geometric and energetic properties of
quasi 1-D wires formed by dangling bonds on Si(100)-H (2 $\times$ 1).
The calculations are performed with density functional theory (DFT).
Infinite wires are found to be insulating and Peierls distorted,
however finite wires develop localized electronic states that
can be of great use for molecular-based devices. The ground state
solution of finite wires does not correspond to a geometrical
distortion but rather to an antiferromagnetic ordering. For
the stability of wires, the presence of abundant H atoms in nearby
Si atoms can be a problem. We have evaluated the energy barriers
for intradimer and intrarow diffusion finding all of them about 1 eV or
larger, even in the case where a H impurity is already sitting on the wire.
These results are encouraging for using dangling-bond wires
in future devices.

\end{abstract}


\maketitle

\section{Introduction}

The study and application of molecular devices entail the creation
of atomic structures both for the active part of the device as well
as for the interface with other devices and the macroscopic
scale.\cite{Joachim} To this aim, the creation of dangling-bond
wires is of particular relevance. Using the scanning tunneling microscope (STM)
several groups have achieved the formation of atomic wires on different
semiconductor surfaces,\cite{Hosaka,hitosugi_jahn-teller_1999,Soukiassian} among which Si (100) has received much
attention~\cite{hitosugi_jahn-teller_1999,Soukiassian} for its properties and its technological use.
Using a completely hydrogen-passivated Si (100) surface, controlled
hydrogen desorption can be achieved with an STM tip.\cite{hitosugi_jahn-teller_1999,Soukiassian} In this way, atomic-size wires of Si atoms can be created
where their properties are very different from the surrounding passivated
Si atoms because they have an open-shell configuration. These dangling-bonds
originate electronic quasi 1-D structures.

The existence of extended structures of aligned dangling bonds has also
led to an intense theoretical activity.\cite{watanabe_theoretical_1996,Doumergue,cho_nature_2002,bowler_small_2000,Bird_SS,Bird,lee_quantum_2009,Kawai,lee_instability_2011} The interest for these systems stems from their pristine conditions
that allow a complete theoretical study together with the unique physical
properties that can be studied. It is well-known that a chain of dangling
bonds leads to extended electronic states that give rise to a partially-filled
1-D band. Its 1-D character
makes it prone to strong correlations and instabilities. Virtually, any
perturbation will destroy the metallicity of the state opening
an electronic band gap at the Fermi level. If the perturbation is due to the electron-vibration
coupling, a Peierls distortion~\cite{Peierls} will set in, and will have
visible consequences in the arrangement of atoms. However, electron-electron
correlations are also present, possibly leading to a Luttinger liquid phase,
and to magnetic correlations. This fundamental interest is enhanced by
the practical consequences of the formation of these dangling-bond wires. In principle,
transport of electrons on an otherwise insulating surface might be
possible.  

Structural experimental data have been obtained
 for finite wires.\cite{hitosugi_jahn-teller_1999,Soukiassian,Lai} The visualization of deformed structures
with the STM~\cite{hitosugi_jahn-teller_1999,Lai} seems to indicate the
existence of strong electron-vibration couplings and the appearance of
a Jahn-Teller effect (the counter-part of the extended-system Peierls
effect). If an extra charge is included to account for transport,
the injected electron propagates
with the wire deformation leading to small polaron
transport.\cite{bowler_small_2000,Bird} Moreover, solitons
have also been predicted as due to the displacement of the domain wall
between adjacent domains of distorted atoms.\cite{Bird} Recently,
solitons have also been revealed in more extended atomic structures on Si
(100).\cite{Lai} Density functional theory (DFT) calculations seem to
indicate that finite systems are not subjected to large electron-vibration
effects because antiferromagnetic ordering counterbalances the atomic
distortions.\cite{Bird_SS,lee_quantum_2009}

In the present article, we perform DFT calculations to explore
the stability of these dangling-bond wires on Si (100) with the
aim of establishing their possible use in future atomic scale devices.
First, we explore infinite wires allowing them to find the total
energy minima as the unit-cell periodicity is changed
 (to allow for different reconstructions) and including spin
polarization. We find that in all these cases gaps develop destroying
the metallicity of the ideal dangling-bond wire in good agreement with
previously reported studies.\cite{cho_nature_2002,lee_instability_2011}
An extra source of electronic localization is the attachment of atomic
impurities at the chemically active dangling-bonds. In particular, the
abundant quantities of H-atoms on the surface, can lead to diffusion
and to the passivation of the wire. To assess the stability of the
wire, we evaluate the reaction pathway to displace one H atom onto
the wire. We repeat the same study for finite wires and study their
properties with system size, including the H-atom diffusion study. Our
calculations show some discrepancy on the behavior with system size of
previous calculations,\cite{lee_quantum_2009} hence we have completed
our calculations with a thorough comparison with two different DFT
implementations, and with different pseudopontential schemes. Our results
indicate that the antiferromagnetic ground state will prevail over the
Jahn-Teller distorted one. However, it is difficult to conclude on the
actual electronic conductance of these systems~\cite{Doumergue,Kepenekian} 
without a careful electron
transport calculation.\cite{Kepenekian} A summary and conclusions
section finalizes the article.

\section{Theoretical method}

{\it Ab initio} calculations were performed within the density-functional theory (DFT) 
framework as implemented in the VASP code.\cite{kresse_efficiency_1996} The PBE form 
generalized gradient approximation was used for the exchange and correlation 
functional.\cite{perdew_generalized_1996} Wave functions 
were expanded using a plane-wave basis set with a cutoff energy of 300~eV. Core
electrons were treated within the projector augmented wave method.\cite{bloechl_projector_1994,kresse_ultrasoft_1999} The surfaces were
modeled using a slab geometry with eight silicon layers in the unit cell.
Eight silicon layers give a good compromise between computational burden
and electronic accuracy. Indeed, while the total energy of the different structures
is largely converged with eight silicon layers, the band structure, and
in particular the bulk gap is only attained at twelve layers. The bottom 
surface was passivated using two hydrogen atoms per silicon one. Different unit cells 
were used for each calculation, as indicated in next sections. The $k$-point sample 
was varied accordingly to the unit cell, and tests were performed to assure its 
convergence. The geometries were optimized until the forces were smaller than 
0.01~eV/\AA. In order to check the consistency of the results, some 
calculations have been repeated with the 
\textsc{Siesta} implementation of DFT,\cite{soler_siesta_2002,artacho_siesta_2008} using Troullier-Martins pseudopotentials,\cite{troullier_efficient_1991}
a double-$\zeta$ polarized basis set to expand the valence wave-functions,\cite{artacho_linearscaling_1999} and an energy cutoff of 200~Ry to set the real space mesh size.

\section{Infinite dangling-bond wire}\label{infinite}
\begin{figure}[ht]
\includegraphics[width=\columnwidth]{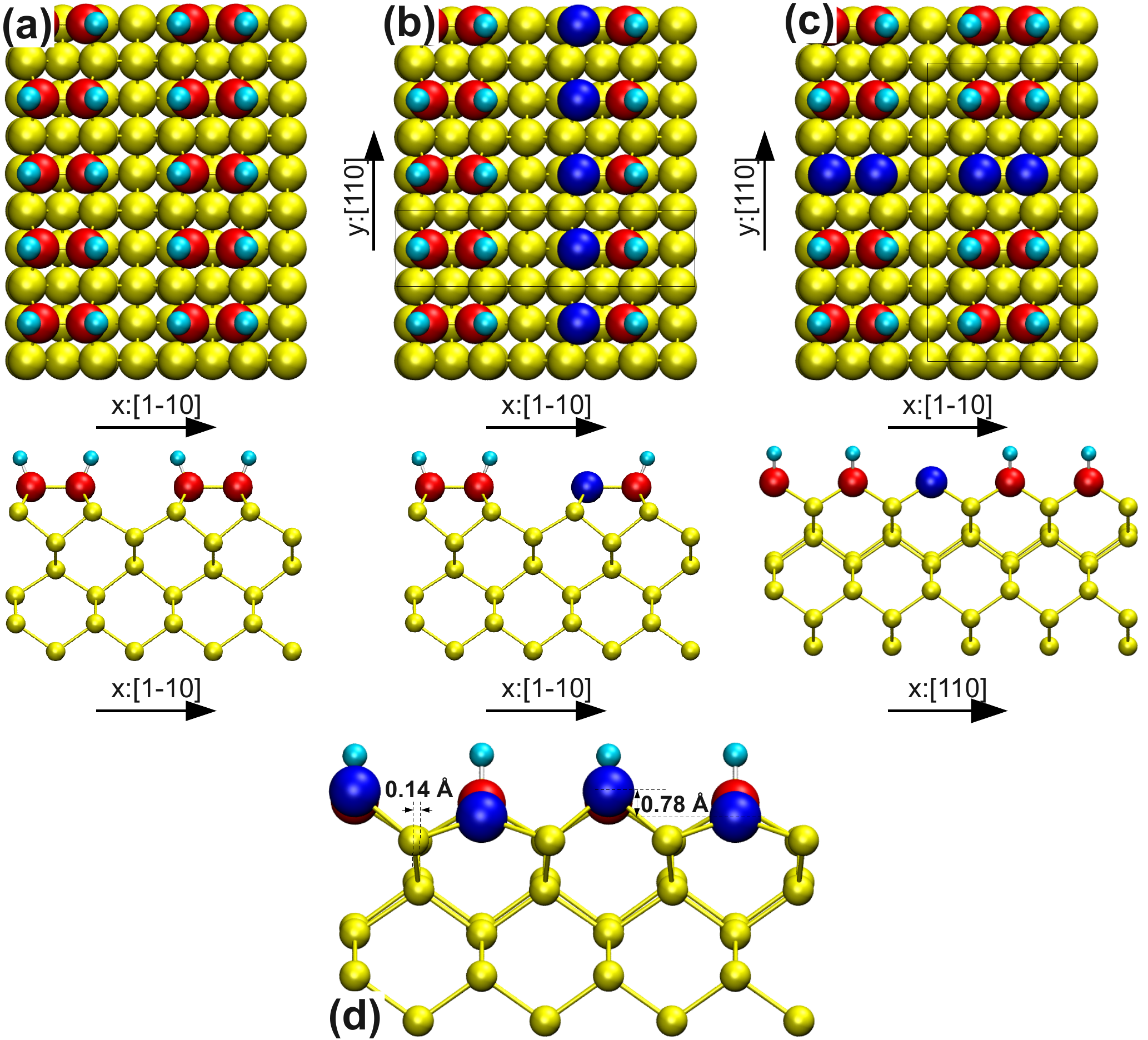}
\caption{\label{geom} (Color online) Cartoons of the completely hydrogenated Si(100)(2x1) surface (a), and DB-wires in y:[110] (b) and x:[1-10] (c) directions are shown. In (d) we show the Peierls-like distortion of the DB-wire in the y-direction when using a (4$\times$2) unit cell. Cyan balls represent hydrogen atoms, while silicon atoms are represented by red (surface dimers), blue (Si with dangling-bonds), and yellow (all the rest) balls.}
\end{figure}
\begin{figure*}[ht]
\includegraphics[width=2\columnwidth]{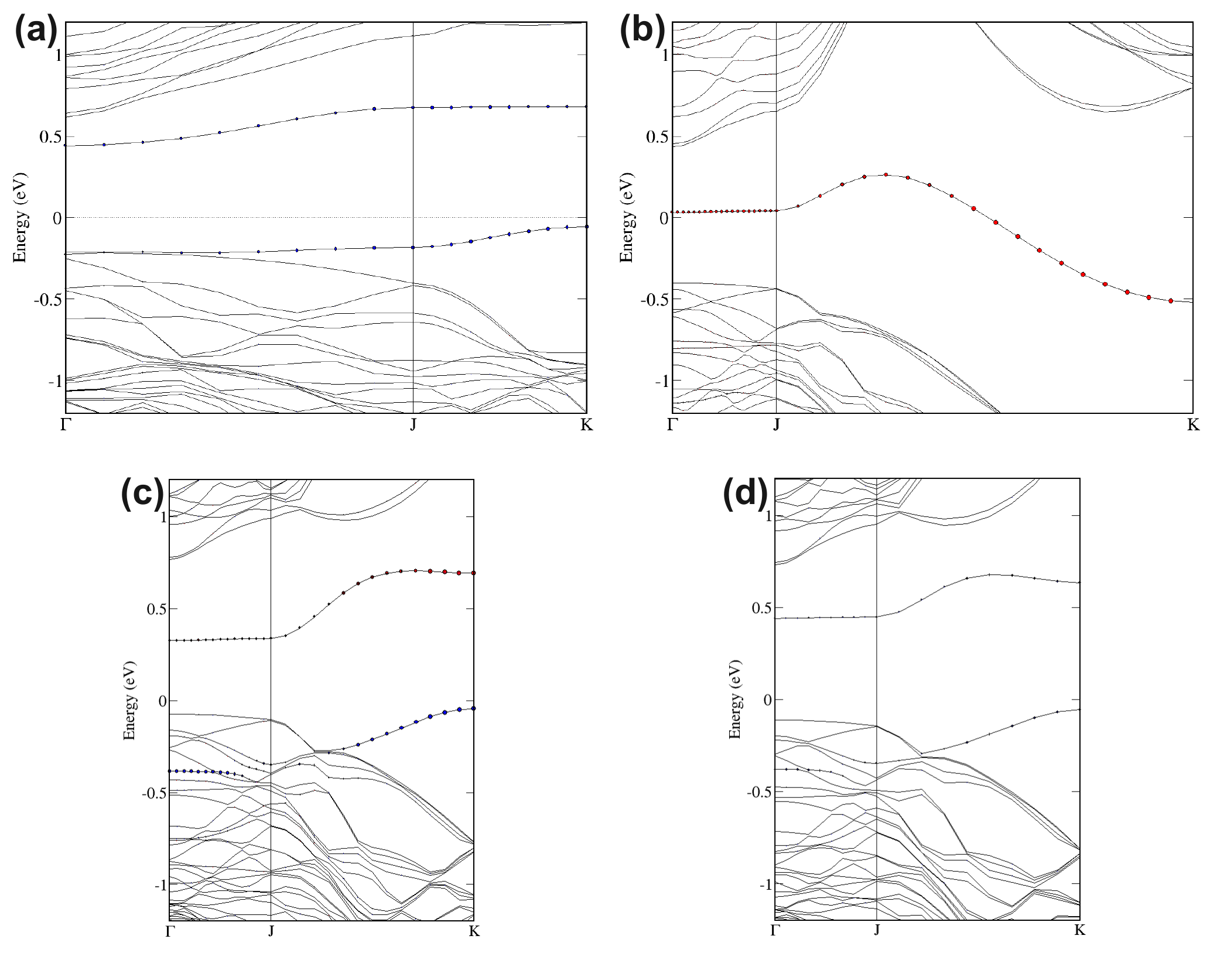}
\caption{\label{bands} (Color online) Electronic bandstructures of several DB-wires: (a) wire in the x-direction; (b) wire in the y-direction using the (4$\times$1) unit cell (ideal case); wire in the y-direction using the (4$\times$2) unit cell in the NM distorted (c) and AFM (d) situations.}
\end{figure*}
We start the discussion with the analysis of the results for an infinite dangling-bond (DB) wire. As shown in Fig.~\ref{geom}a, the completely hydrogenated Si(100) forms a (2$\times$1) reconstruction by the creation of dimer rows along the y:[110] direction. A DB-wire can be constructed along two perpendicular directions, as we remove a line of hydrogens along the dimer rows direction (y), or perpendicular to it (x:[1-10]) (Figs.~\ref{geom}b and c, respectively). To study the DB-wire along the x-direction, we use the (2$\times$4) unit cell shown in Fig.~\ref{geom}c. The band structure of the system is shown in Fig.~\ref{bands}a. With dots we show the bands with weight from the DB silicons. We can observe that two bands appear in the bulk energy gap, and they mainly come from the DB-silicons. The gap is reduced with respect to the completely passivated surface, but we still obtain 0.66~eV for the direct band gap, and 0.50~eV for the indirect one. 
\begin{table*}[ht]
\centering
\caption[] {\label{tab1} Calculated height difference $\Delta_{up-dw}$(\AA) between up and down atoms in the distorted DB wire, direct and indirect band gap energy E$_g$ (eV), and energy gain $\Delta E$ (meV/DB) with respect to the ideal NM DB wire. Several previous results are included.}
\begin{ruledtabular}
\begin{tabular}{clcccc}
         &  &   $\Delta d_{up-dw}$(\AA) &  E$_g$(direct)(eV) & E$_g$(indirect)(eV) &  $\Delta E$(meV/DB) \\
        \hline
        NM distorted &(Watanabe {\it et al}) \footnotemark[1] & 0.16 & 0.126 & 0.025 & 7    \\ 
         &(Hitosugi {\it et al}) \footnotemark[2] & 0.50 &       &       & 47   \\
         &(Bowler {\it et al})   \footnotemark[3] & 0.67 &       &       & 105  \\
         &(Cho {\it et al})      \footnotemark[4] & 0.74 & 0.68  & 0.36  & 43   \\
         &(Lee {\it et al})      \footnotemark[5] & 0.78 &       & 0.39  & 49   \\
         & VASP                 & 0.78 & 0.40  & 0.36  & 56   \\
         & {\sc Siesta}               & 0.81 & 0.45  & 0.44  & 61   \\
        \hline
        AFM &(Lee {\it et al})               \footnotemark[5] & 0.00 &       & 0.63  & 57   \\
         & VASP                          & 0.00 & 0.55  & 0.50  & 41   \\
         & {\sc Siesta}                        & 0.00 & 0.63  & 0.60  & 56   \\
\end{tabular}
\end{ruledtabular}
\footnotetext[1]{Ref.~\onlinecite{watanabe_theoretical_1996}.}
\footnotetext[2]{Ref.~\onlinecite{hitosugi_jahn-teller_1999}.}
\footnotetext[3]{Ref.~\onlinecite{bowler_small_2000}.}
\footnotetext[4]{Ref.~\onlinecite{cho_nature_2002}.}
\footnotetext[5]{Ref.~\onlinecite{lee_quantum_2009}.}
\centering
\end{table*}
In order to study the wire along the y-direction, we start by
using a (4$\times$1) unit cell (Fig.~\ref{geom}b). In this case the
situation qualitatively changes, and we observe a double degenerate
metallic band crossing the Fermi energy, coming from the DB-silicons
(Fig.~\ref{bands}b). The peculiar shape of the dispersion curve of this 
metallic band can be explained by analyzing the competition between $p_z$ and 
$s$ electronic coupling considering the top surface and the sub surface Si 
atoms with a negligible through space direct electronic coupling between the 
Si top atoms.\cite{Kawai,Kepenekian}

However, the stabilization of the metallic band is only achieved by
the constriction imposed by the (4$\times$1) unit cell, in which all the
atoms belonging to the DB-wire are equivalent. This ideal case can
be destabilized by several mechanisms, like spin polarization or the
formation of a charge density wave along the wire direction, which could
give rise to a Peierls-like
situation,\cite{bowler_small_2000,lee_quantum_2009} To check these
possibilities we have doubled our unit cell in the y-direction to obtain a
(4$\times$2) cell in which we have two inequivalent DB-silicon atoms.

First, we have calculated the frequencies of the system to check its
stability, finding an imaginary frequency which reveals that the system
is in a saddle point. In the mode corresponding to this frequency
one of the atoms in the DB-wire goes upwards, while the other goes
downwards. If we relax the new unit cell considering this possibility,
we actually find a more stable geometry in which the two atoms in the
new distorted wire are separated by 0.78~\AA~ in the vertical direction
(Fig.~\ref{geom}d). This new configuration is more stable than the
ideal one by 56~meV/DB. The new bandstructure of the system can be seen
in Fig.~\ref{bands}c. The metallic band of the ideal DB-wire is now
split into two bands, opening a direct (indirect) band gap of 0.40~eV
(0.36~eV). The new occupied band comes mainly from the DB-silicon in
the upper position, which is stabilized by losing 0.11 electrons, while
the unoccupied band states are mainly coming from the Si atom in the down
position, which receives 0.17~e$^-$. Besides the zig-zag distortion of the
atoms in the DB-wire, the underneath Si atoms undergo a slight dimerization
(Fig.~\ref{geom}d). We have not found any more complicated configuration
which could exist when increasing the unit cell size. Actually, the
distorted solution resembles the dimer buckling which appears in the
non-hydrogenated Si(100) surface, and therefore we believe it is the
most stable configuration that can be found by geometrical relaxation
of the system.

\subsection{Magnetic ground state}

Another possibility to destabilize the ideal DB-wire is to
have a spin-polarization of the system. We have used the (4$\times$2)
unit cell, considering ferromagnetic (FM) and antiferromagnetic (AFM)
couplings between the atoms in the DB wire. The FM case is less stable
than the ideal DB-wire by 27~meV/DB, but the AFM one is more stable
by 41~meV/DB. Each Si atom in the DB-wire has a magnetic moment of
0.50$\mu_B$, and couples antiferromagneticaly to its neighbors in the
wire. In Fig.~\ref{bands}d we show the band structure of the system. It
is similar to the distorted case (Fig.~\ref{bands}c), with a direct
(indirect) band gap of 0.55~eV (0.50~eV). In this case the occupied
band has contributions from the majority spin of one silicon atom in
the DB-wire, and the minority of the adjacent, while in the unoccupied
band we find the minority spin of the first atom, and the majority of
the second. It is interesting to note that both mechanisms of  wire
destabilization are not compatible. The relaxed geometry of the AFM
wire is essentially identical to the ideal case in a (4$\times$1)
unit cell. Inducing a zig-zag distortion within the wire progressively
decreases the magnetic moment of the atoms, until it completely disappears
with a vertical separation of the atoms of 0.50~\AA. A spin polarization
solution can not be stabilized with the relaxed distorted geometry.

The results presented so far have been obtain by using the VASP code. They
have also repeated the calculations with the {\sc Siesta} implementation
of DFT, obtaining the same qualitative results. The quantitative data are
presented in Table~\ref{tab1}.  Our results confirm previous experimental
and theoretical data. Watanabe {\it et al}\cite{watanabe_theoretical_1996}
predicted the Peierls instability in this system. Hitosugi {\it et
al}\cite{hitosugi_jahn-teller_1999} experimentally constructed finite
DB-wires, and observed a lattice distortion which was described by
a Jahn-Teller mechanism. Other theoretical groups have also performed
calculations on the system. All the previous results, together with ours,
are presented in Tab.~\ref{tab1}.

In all cases the most stable configuration has a band gap, which size
depends on the details of the calculation. In ours (VASP and {\sc
Siesta}), the distorted case is more stable than the AFM solution by 15
and 5 meV/DB, respectively, while Lee {\it et al}\cite{lee_quantum_2009}
find the opposite, with the AFM configuration being more stable by
8~meV/DB. The small energy difference between the solutions of the
order of meV reveals that both are competing and, in practice, other
effects like temperature would determine which case could be experimentally observed.
To analyze the nature of this competition, Lee and
Cho~\cite{lee_instability_2011} have studied the DB-wires in C and Ge,
and they have compared them to the Si case. They have shown that the
on-site electron-electron interactions, modelized by a Hubbard-$U$ term, 
are the origin of the AFM coupling, and their strength decreases as we
go down in the periodic table from C to Ge. The smaller atoms have the larger
$U$. This effect has also been reported in the magnetization of SiC surfaces
when an impurity is adsorbed on a DB, showing different behavior
on whether the impurity adsorbs on a C atom or a Si one due to the larger
$U$ of the C atom.\cite{Adrien}

The distorted NM case is due to the electron-lattice coupling,
which increases following the same sequence. Thus, while for C the AFM
case is clearly the most stable, for Ge the distorted configuration
is lower in energy. Si is in an intermediate case where a competition
between both situations is observed.

\subsection{Mechanisms driving the instabilities}

The instabilities found for the 1-D DB wire have the common underlying principle
that the electronic energy of the system is greatly reduced if a
gap opens in the ideal metallic DB wire. This gap opening is driven either
by the geometrical distortion of the wire also known as Peierls instability~\cite{Peierls}
or by the spin polarization of the electronic states.\cite{Anderson}

In the Peierls instability, the geometrical distortion increases the system's energy,
but the gain in electronic energy overcompensates the ionic energy loss. This can
be estimated by separating the global energy gain $\Delta E_{Peierls}$ in
an electronic, $\Delta E_{elec}$ and an ionic. $\Delta E_{ion}$ components. The
first component can be expressed by the difference between the
electronic energy of the distorted
and ideal systems. This energy  is given by the sum over all occupied electronic
states, hence
\begin{equation}
\Delta E_{elec} = \int_{occ} E (DOS_{Peierls} (E) - DOS_{ideal} (E)) dE,
\label{deltaelecpeierls}
\end{equation}
where $DOS$ refers to the electronic density of states. This quantity
is easily evaluated and gives $\Delta E_{elec} \approx -0.4$~eV.

From Table~\ref{tab1} we know that the global energy gain $\Delta E_{Peierls}$
for two DB is
$\approx -0.12$ eV. Then, the energy used to distort two DB in the wire is then $\Delta E_{ion}
\approx 0.28$ eV.

Similarly, spin polarization is favorable only in certain situations.\cite{Anderson} Typically, the gain
produced by the opening of the gap has to overcome the tendency to pair spins in electronic
states. A Stoner-like criterion~\cite{Anderson} for local electronic states shows that spin polarization
is energetically possible if the intra-atomic Coulomb repulsion, $U$, is larger that the energy
spread of electronic states at the Fermi energy. This energy spread is basically the inverse of
the $DOS$ at the Fermi energy, $E_F$. Hence, spin polarization becomes energetically favorable
when
\begin{equation}
U  \times DOS (E_F) \gg 1.
\end{equation}
Hence, the larger $U$ is the more likely it is to find an AFM ground state 
for the DB wire. This has been shown to be the case for C(100) DB wires.
\cite{lee_instability_2011}
Diamond surfaces are actually prone to developing AFM
groundstates~\cite{Yndurain} due to the large valuer of $U$ and the quasi 1-D 
arrangement of atoms on certain
surfaces. Indeed, in Ref.~\onlinecite{lee_instability_2011} the computed 
values of $U$ are 3.45 eV for C DB wires,
while only 1.71 eV for the Si ones. Taking $DOS (E_F) \approx 1.4$
eV$^{-1}$, from the metallic band of the ideal wire, we obtain $U \times DOS 
(E_F) \approx 2.4$. Despite being larger
than 1, this value is not very big, showing the minor tendency of the
Si DB wires to become spin polarized. This argument is closely related
to the fact that the energy of the AFM solution is slightly larger than
the Peierls one, Tab.~\ref{tab1}.

\begin{figure}[ht]
\includegraphics[width=\columnwidth]{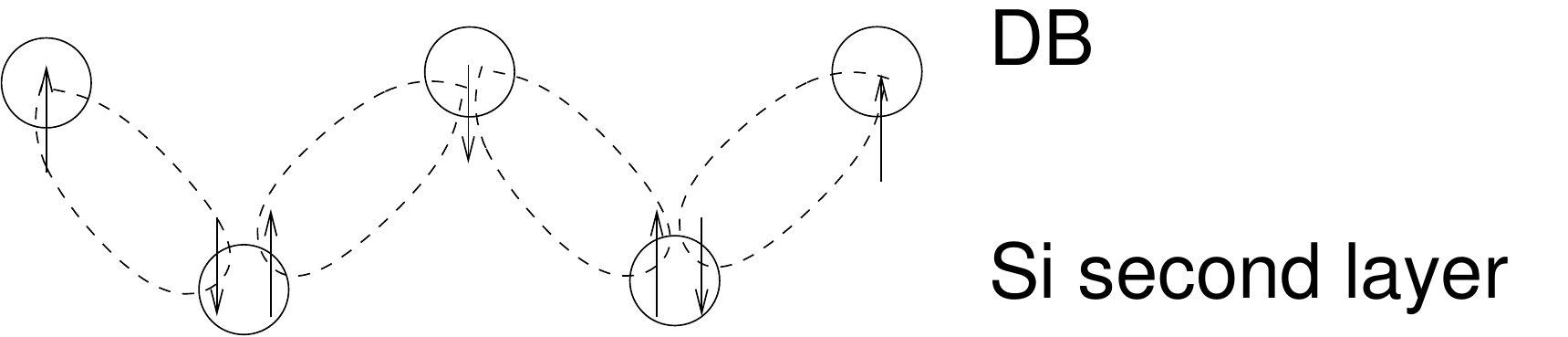} 
\caption{
\label{scheme}
Scheme for the superexchange mechanism leading to the anti-ferromagnetic (AFM)
ground state of the DB wire. The second-layer
atoms are not polarized. The covalent bond with
the DB atoms is represented by a dashed line. The spin pairing in these
covalent bonds force the AFM  ordering of the DB spins.
}
\end{figure}

The AFM solution is driven by superexchange due to the second layer Si atoms, 
ruling out ferromagnetic solutions for the ground state of the DB wire. It is 
easy to visualize the superexchange mechanism in these wires since the second 
layer has negligible spin polarization and it is covalently bound
to the spin polarized DB orbitals. Hence, spins pair between the DB and the 
second-layer atom, while the second-layer atom is shared between to 
neighboring DB that forcedly have opposite spins to keep the second-layer atom 
unpolarized. See the scheme of Fig.~\ref{scheme}.

\subsection{Stability of the wire vs. hydrogen diffusion}\label{diffusion}

A crucial aspect to study before considering DB-wires for any
technological application is the chemical stability of the system. One of the
mechanisms which could destabilize the DB-wires is the surface diffusion of
hydrogen atoms, which by bonding with a silicon DB along the wire would
eliminate one DB, modifying the wire electronic transport 
properties.\cite{Kawai}
The new H atom could
come from several sources. Let us first consider how the diffusion of
hydrogen from the adjacent Si in the dimer would affect the stability
of the wire in the y-direction. 

\begin{figure}[ht]
\includegraphics[width=\columnwidth]{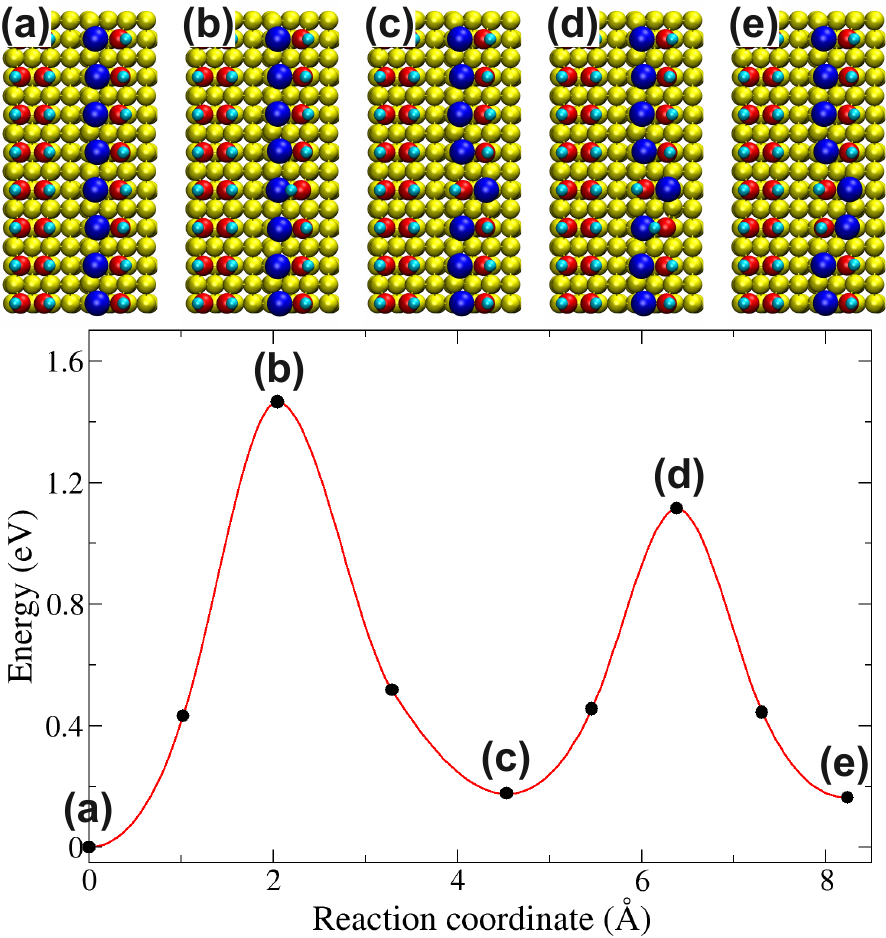} 
\caption{
\label{diff}
(Color online) Reaction path for the diffusion of one and two hydrogens to the DB-wire
from the adjacent dimer, as shown in the upper panels. The color code
is the same of Fig.~\ref{geom}.} 
\end{figure}

To perform the study we have used a
($4 \times 8$) unit cell, with enough atoms in the DB-wire to consider the H
impurity as isolated along the wire's direction. 
First, we have calculated the energy difference of
having one H moving to the DB-wire from the closest position, as shown
in panel (c) of Fig.~\ref{diff}. The final relaxed energy is 0.18~eV
higher in energy than the pristine wire with all H aligned on one side
of the row and the other side forming the DB wire.  Creating
one H vacancy by displacing it to the DB wire 
is not energetically favorable. We
have also calculated the energy barrier of the process using the
nudge elastic band (NEB) method.\cite{sheppard_optimization_2008}
The predicted reaction path is shown in the left part of Fig.~\ref{diff}. The
transition state is shown in Fig.~\ref{diff}~(b), and it corresponds
to a large barrier of 1.47~eV. Therefore, in principle the DB-wire
would be stable vs. this kind of processes. 

Our theoretical barrier of 1.47~eV is in excellent agreement with
the measured value of 1.46~eV.\cite{Hoefer2010} However, the experimental
situation does not correspond to the same situation computed here. Here,
we are interested in the intradimer diffusion when an infinite DB wire is
form. In the experiment, they rather have a two-DB wire (formed by the
desorption of a H$_2$ molecule after a laser pulse) and have intradimer
diffusion. This situation corresponds to the one of finite wires
that we will see in Section~\ref{finiteH}.

In the case that a
hydrogen atom has already jumped to the DB-wire, we may ask which would be the
probability for a second H to jump, as shown in Fig.~\ref{diff}~(d). This
final configuration is 0.01~eV lower in energy than the previous one, and the
energy barrier is also lowered to 0.94~eV, as shown in the right part of
Fig.~\ref{diff}. Hence after the first hydrogen has jumped, it is easier for
a second hydrogen to do the same, than for the first one to jump back. 
However, the lowest total energy is found in the later situation 
showing that the pristine wire is more stable.

Alternatively,
after the first hydrogen jumps it could diffuse along the wire, jumping to
the next DB. Our calculations yield that this
intrarow diffusion would decrease the final total energy in 0.23~eV. After
the pristine wire, the lower-energy state corresponds then to
an H atom that has diffused across the dimer and along the wire.
The barrier
associated to the intrarow diffusion is 1.44~eV. This result also shows
that it would not be easy for a H atom to diffuse along a DB wire.
Even if an extra H atom 
is placed in the DB wire coming from another source
the intrarow diffusion is large, amounting to
1.36~eV,
similar in value to the previous intrarow-diffusion barrier.

Experimentally, the reported values for intrarow-diffusion barriers 
are about 1.68 - 1.75 eV.\cite{Owen,Hill} These are $\sim 0.2$-eV
higher than our computed value. Discrepancies are probably due
to the different H coverage of the two systems compared here. Indeed,
thorough computations~\cite{Pollmann} with different H coverages yield
a $0.25$-eV difference between their computed
high H-coverage and low coverage limits.

In conclusion, our stability studies show that the diffusion of hydrogen to the wire or along it is not a very favorable process, with associated energy barriers between 1 and 1.5~eV. Therefore a DB wire can be considered as a quite stable structure when considering those lateral and on line H surface diffusions. 

\section{Finite wires}\label{finite}
\begin{figure}[ht]
\includegraphics[width=\columnwidth]{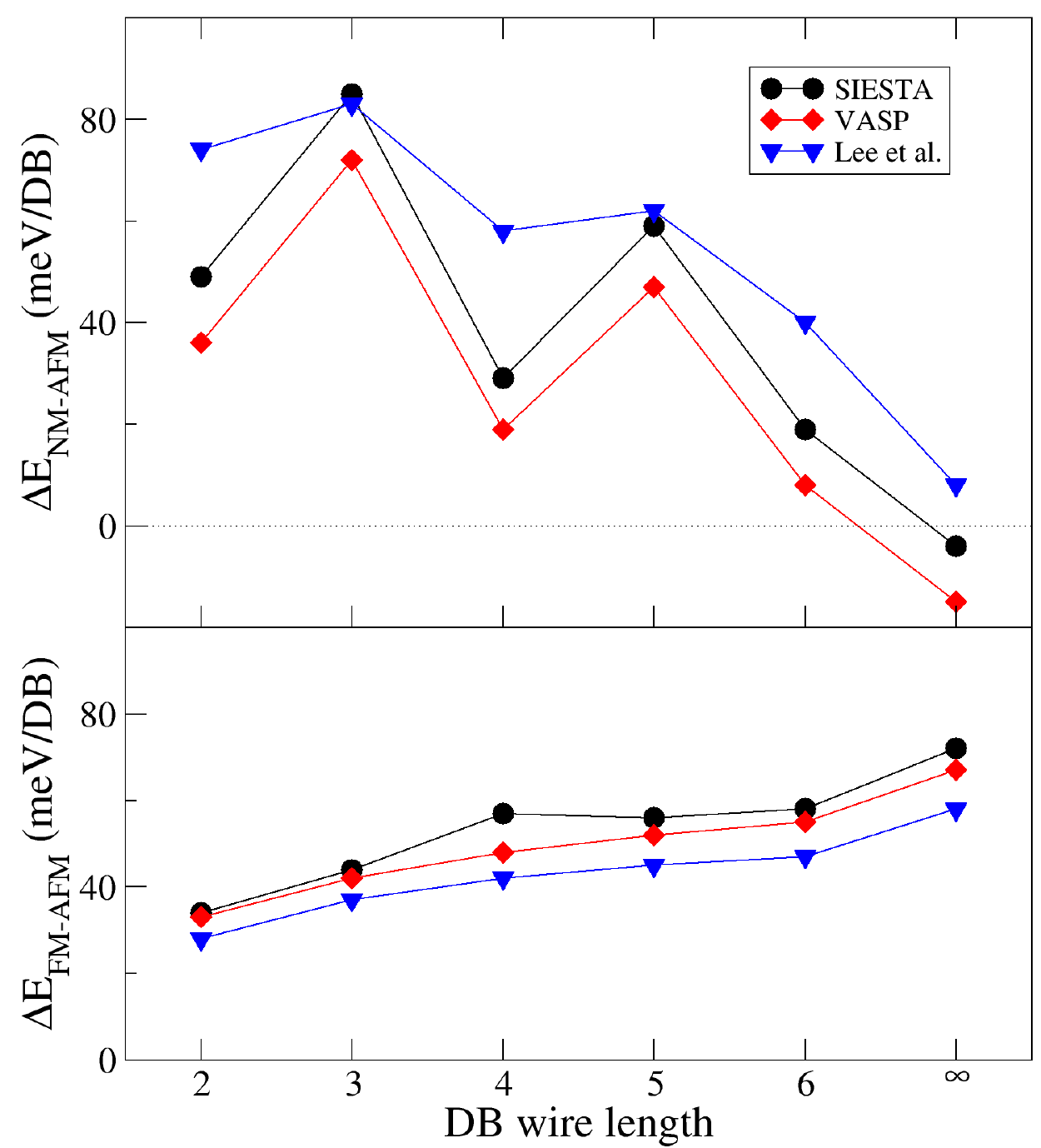}
\caption{\label{finitos} (Color online) Energy differences (in meV/DB) between NM distorted and AFM configurations (upper panel), and between FM and AFM configurations (lower panel). Results for {\sc Siesta} (circles), VASP (diamonds), and from Lee~{\it et al}\cite{lee_quantum_2009} (triangles) are shown.}
\end{figure}
In previous sections we have investigated the properties of infinite
DB-wires, which can represent the properties of very extended
wires. However, in practice we will deal with shorter DB wires in which
finite-size effects can be important. Experimentally, Hitosugi~{\it et
al}\cite{hitosugi_jahn-teller_1999} constructed wires ranging from
2 to 13 DB. 

In order to investigate which are the properties of finite
wires and how they change as their size increases, we have considered
wires from 2 to 6 DB in a (4$\times$8) unit cell. The results are summarized
in Fig.~\ref{finitos}, and compared with previous results from Lee~{\it
et al}.\cite{lee_quantum_2009} 

In general, we observe that the magnetic
solutions are favored for all the finite wire sizes. The AFM solution is
the most stable in all cases, although the NM-distorted case competes
more in energy as the DB-wire length is increased. Even the FM case,
which was less stable than the ideal case for the infinite wires, is
preferred to the distorted case for the smaller cases, up to 3-DB in
our {\sc Siesta} and VASP calculations, and up to 5-DB in the ones from
Lee~{\it et al}.\cite{lee_quantum_2009} 

VASP and {\sc Siesta} methods give a similar behavior,
with the results shifted by almost a constant amount. This systematic
discrepancy (less than 10 meV) is most likely due to the different treatment
of the core electrons in both methods: PAW 
potentials\cite{bloechl_projector_1994} for
VASP and norm-conserving pseudopotentials\cite{troullier_efficient_1991}
for {\sc Siesta}. 

In the upper panel of Fig.~\ref{finitos} a pronounced
even-odd oscillation is found. It can be traced back to oscillations
in the stability of NM-distorted wires. Their origin was assigned
by Lee~{\it et al} to the presence of a half-filled electronic state for the 
odd wires,\cite{lee_quantum_2009} which destabilize them compared to the even
ones. Actually, in our calculations we observe that the NM-distorted solution
for the odd wires is a local minimum only if the system is forced to be
non magnetic. If spin polarization is allowed, the NM-distorted solution
is not longer stable and it evolves until converging to the AFM solution. 

The results from Lee~{\it et al} agree well with ours for the magnetic cases,
but there is a discrepancy for the NM-distorted situations. Here, our 
VASP and {\sc Siesta} calculations predict a higher stability for the 
even-numbered wires, which make the even-odd oscillations in the upper panel 
of Fig.~\ref{finitos} more pronounced in our case. We assign this discrepancy 
to the need of a thorough atomic relaxation. Curiously enough, the agreement
is better for the spurious NM-distorted solutions in the odd-numbered wires.

Due to the different number of atoms in the finite wires, even-numbered 
NM-distorted wires
can have a lone Si atom in the second layer that plays the
role of a soliton wall.\cite{Bird} 
The average distortion increases as we
increase the length of the wire, reaching its maximum for the infinite
one. However, the convergence is fast, and for the 6-DB the values are
very close to the infinite-wire ones. 
In the AFM configuration the values for the local magnetic moments are the
same to the infinite case (0.50$\mu_B$) except at the edges, where they
increase to 0.56$\mu_B$. 

In general, we can conclude that, for the finite wires, the magnetic cases 
are more favorable than the distortion of the lattice, probably because of 
the increased electronic confinement of the system. Indeed the magnetic
solutions are the only ones for wires with an odd number of DBs.

 \begin{figure}[ht] 
\includegraphics[width=\columnwidth]{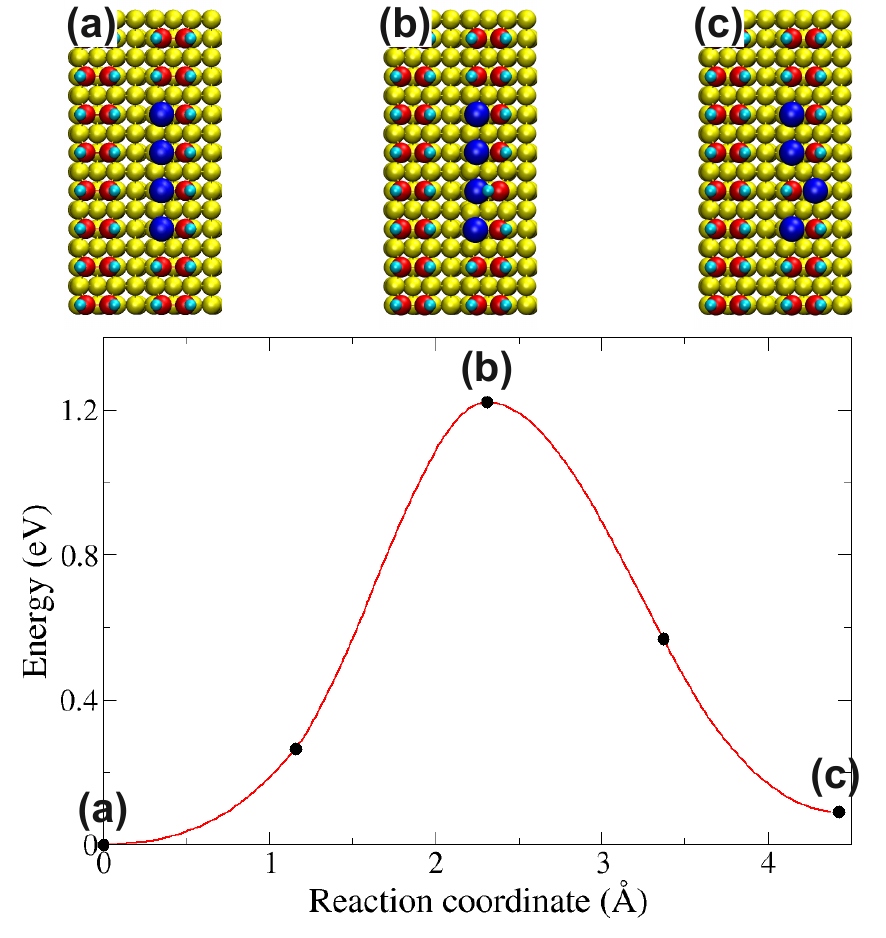}
\caption{\label{diff_fin} (Color online) Reaction path for the intradimer diffusion of one hydrogen
in the 4-DB wire, as shown in the upper
panels. The color code is the same of Fig.~\ref{geom}.} 
\end{figure} %

\subsection{Finite-wire Stability vs. hydrogen diffusion}
\label{finiteH}

As we did for the infinite wires, it is worth to study the stability
of the finite wires. In this case we will start by studying the possibility of an H
atom to intradimer diffuse in the 4-DB wire, as shown
in the upper panels of Fig.~\ref{diff_fin}. Our calculations show that,
while the final situation is just slightly higher in energy (0.09~eV),
the associated energy barrier is of 1.22~eV, similar to the cases studied
in Section~\ref{diffusion}.

Regarding the intrarow diffusion, when a H atom diffuses from the edge of the 4-DB wire
to the inside, we obtain a barrier of 1.54~eV, also similar to the corresponding value 
for the infinite wire.

These results are in good agreement with previous calculations
at different levels of H coverage on Si (001), where a systematic
larger intrarow barrier is found compared to the intradimer
one.\cite{Pollmann,Owen} Diffusion between rows is about one eV larger
than the intrarow diffusion barrier~\cite{Hoefer2010,Pollmann,Owen}
and we can effectively neglect it as a threat to the DB-wire stability.

Consequently, the finite wires can also be
considered as stable following this criterion.

\section{Summary and conclusions}

The present article shows the properties of quasi 1-D wires formed
by lines of dangling-bonds (DB) created by the removal of H atoms
from a monohydride Si (100) surface. In particular, emphasis is put
on their stability and their capacity to be used in molecular-based
devices.

The ideal infinite DB wire presents a narrow metallic band and it is
thus subjected to instabilities. Electron-vibration coupling
opens a gap while inducing a Peierls distortion of the wire. Electron-electron
interactions leads to different instabilities, in particular ferromagnetic
and antiferromagnetic solutions that also open a band gap. Hence, realistic
infinite DB wires are not metallic. According to our two calculations
with different DFT-based methods, the ground state of the infinite wire
corresponds to the Peierls solutions, with an important degree of geometrical
distortion and a sizeable band gap, approaching the bulk gap.

Wires can be considerably perturbed if H atoms move away from
neighboring passivated bonds. Our computations show that the
lower diffusion barriers correspond to intradimer diffusion with
the value of 1.47 eV in excellent agreement with experimental
measurements.\cite{Hoefer2010} Once a H has diffused into
the wire, the barrier to diffuse a second neighboring H atom
goes down to 0.94 eV. Despite this reduction, the large activation
barriers seem to warranty the stability of DB wires.

Actual devices will rather include finite wires. Finite effects
change the ground state of the wires from the non-magnetic distorted
solution to the antiferromagnetic one. The short wires behave like molecular
systems with discrete electronic levels that can be of great
help to shape the electronic transparency of those DB wires. 

Finite wires have a larger number of neighboring H atoms. But our study
shows that the intrarow-diffusion barrier is still very large in these
systems (1.54 eV) and the bottle neck for H diffusion is the intradimer
one. Hence, we expect DB wires to be stable against H contamination. 

Given the robustness
of these systems against other perturbations and contamination by
neighboring atoms, we expect short DB wires to gain importance in the
creation of atomic scale devices.\cite{Kawai_2012}

\acknowledgments 
This work has been supported by the ICT-FET European Union Integrated Project
{\em AtMol} (www.atmol.eu).

\bibliographystyle{apsrev}

\end{document}